\documentclass[twocolumn]{jpsj2}
\usepackage{graphicx}
\usepackage{dcolumn}

\newcommand{\bm}[1]{\mbox{\boldmath{$#1$}}}

\title{Entanglement Entropy of 1D Gapped Spin Chains}%

\author{
Takaaki \textsc{Hirano}$^1$\thanks{E-mail: hirano@pothos.t.u-tokyo.ac.jp} and 
Yasuhiro \textsc{Hatsugai}$^{1,2}$\thanks{E-mail:hatsugai@pothos.t.u-tokyo.ac.jp;
$ $hatsugai@sakura.cc.tsukuba.ac.jp}
}

\inst{$^1$ Department of Applied Physics, University of Tokyo, 7-3-1 Hongo,
Bunkyo-ku, Tokyo 113-8656, Japan\\
$^2$ Institute of Physics, University of Tsukuba, 1-1-1 Tennodai
Tukuba, Ibaraki 305-8571 Japan}

\kword{Topological orders, Quantum phase transition, Entanglement entropy, Spin chains, Haldane gap, Edge states}

\abst{
We investigate the entanglement entropy (EE) of gapped 
$S=1$ and $S=1/2$ spin chains with dimerization.
 We find that the effective boundary degrees of freedom as edge states
contribute significantly to the EE.
For the $S=1/2$  dimerized Heisenberg chain, the EE
 of the sufficiently long chain is essentially explained by the localized $S=1/2$
 effective spins on the boundaries.
As for $S=1$, the effective spins are also $S=1/2$ causing a Kennedy
 triplet that yields a lower bound for the EE.
In this case, the residual entanglement reduces substantially by 
a continuous deformation of the Heisenberg model to that of the AKLT Hamiltonian.
}

\begin{document}
\maketitle

\section{Introduction}
The concept of entanglement in quantum mechanics has significant importance in
quantum information technologies, where quantum resources are used for
processing information in novel ways. 
Recently, it has been clarified that the entanglement of a quantum state is
one of the most important properties not only in quantum information
science but also in condensed matter physics.
Entanglement entropy (EE) measures the entanglement of a single quantum
state, which is defined as the von Neumann entropy of a reduced
 density matrix of a subsystem.
EE has been applied to a condensed matter system in order
      to understand quantum phases.
There are several
works\cite{Osterloh,Osborne,Vidal,Refael} that
have reported the EE near or at a quantum critical point.
As for the system size dependence, the EE of gapped spin chains
saturates, although that of
critical spin chains shows a logarithmic divergence.
Recently, the EE was used to detect the quantum dimension, which is 
a property of the topological order\cite{TEE1,TEE2,Ryu}.
The topological order is a new order that cannot be characterized by
a classical local
order parameter and a symmetry breaking by the conventional
Ginzburg-Landau theory\cite{TOrder}.
Recently, one of the authors proposed the use of quantum objects such as
quantized Berry phases to define the local order parameters that do not
require classical symmetry breaking since they are gauge-dependent
quantities unlike the classical ones
\cite{HatsugaiOrder1,HatsugaiOrder2,HatsugaiOrder3,Maruyama}.

It is also interesting to study the EE from the
viewpoint of bulk-edge correspondence.
Although the EE is a physical quantity of the bulk, 
there is a relation between the EE and the edge states
that appear in a topologically nontrivial system with boundaries.
There are several works on the bulk-edge correspondence in many
kinds of physical systems such as the 2D quantum Hall
effect\cite{LaughlinEdge,HatsugaiEdge},
polyacetylene\cite{SSH,RyuEdge},
and the Haldane spin chain\cite{KennedyEdge,Hagiwara}.
The EE and 
 the Berry phase of a topologically nontrivial system are closely related through the edge states\cite{Ryu}.

In this paper, we investigate the relation between the saturated EE in
 a gapped system and the corresponding edge state through the 
bulk-edge correspondence. 
The systems chosen are three different types of gapped quantum spin
chains: (i) the $S=1/2$ dimerized Heisenberg chain, (ii) the $S=1$ XXZ
Hamiltonian with the on-site anisotropic term
$D\sum_{i=1}^N\left(S_i^z\right)^2$, and (iii) the Affleck Kennedy Lieb
Tasaki (AKLT) model\cite{AKLT}.

The $S=1$ XXZ chain near the Heisenberg point is a typical Haldane
spin chain, which has a finite gap in an appropriate range of
the parameters. 
There are several other phases such as the ``large-$D$ phase'' and the ``N\'{e}el phase.''
They are characterized by the hidden
$\bm{Z}_2\times\bm{Z}_2$ symmetry breaking, which is nonlocal\cite{Z2Z2-1,Z2Z2-2}.
The string order parameter is a nonlocal order parameter\cite{Nijs};
therefore, it is
not a conventional classical order parameter. 
We see that the EE reflects these kinds of hidden symmetry breaking,
which relates to the structure of the edge states.

\section{EE of gapped system and edge state}
The EE (the von-Neumann entropy) used in this paper is defined as
follows:
\begin{eqnarray}
{\cal S} &=& -\langle\log \hat{\rho}_{A}\rangle =-Tr_{A}[\hat{\rho}_{A} \log \hat{\rho}_{A}  ],
\end{eqnarray}
$\hat{\rho} _{A} = Tr_B[ \hat{\rho} ]$, 
where $\hat{\rho}$ is a density matrix of the pure state $|\psi\rangle$.
We assume that the system consists of subsystems A and B.
The reduced density matrix $\hat{\rho}_A$ is obtained by tracing out the
degrees of freedom in subsystem B from the total density matrix $\hat{\rho}$. 
The EE quantifies the information about how much the state is
 entangled between the subsystems A and B. We consider $| \psi
 \rangle$ to be the ground state (GS), assuming that it is unique.

We divide a one-dimensional chain into 
subsystems A and B (Fig.\ref{fig:PBC-OBC}) that denote the upper and
lower parts, respectively. 
Since taking a partial trace over a spatially separated subsystem
results in effective boundaries, it induces effective localized degrees of freedom
as edge states.
Then, we evaluate the EE as ${\cal S}\sim{\cal S}_{bulk}+{\cal
S}_{edge}$ where ${\cal S}_{bulk}$ and ${\cal S}_{edge}$ denote the
contribution from the bulk and edge, respectively.
This is because the degrees of freedom originate not only from the edges but
also from the bulk part of the system.
We hypothesize ${\cal S}_{bulk}\ge 0$ and, we obtain the relation 
${\cal S}\ge{\cal S}_{edge}$, which means
that the EE has lower bound of ${\cal S}_{edge}$.
The contribution from the edges becomes
important in a gapped system,
while that from the bulk is essential in a gapless system
because of the infinite correlation length\cite{Ryu,Katsura}.
\begin{figure}[!tb]
\begin{center}
\includegraphics[width=8cm]{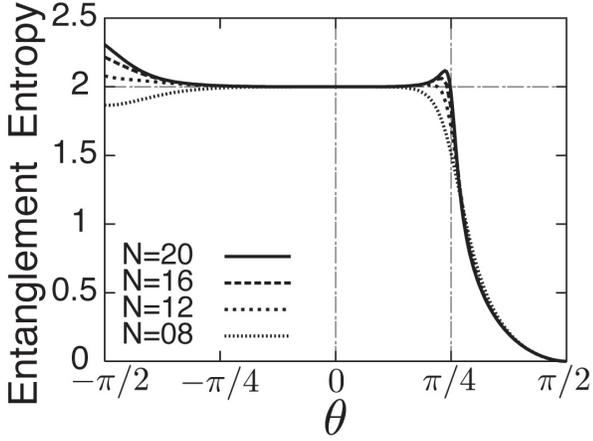}
\end{center}
\caption{The EE of the periodic dimerized Heisenberg
 chains. $J_1=\sin\theta$ and $J_2=\cos\theta$, and the system size is
 $N=8,12,16,20$. The base of the logarithm is $2$.}
\label{fig:dimerf}
\end{figure}
\begin{figure}[!tb]
\begin{center}
 \includegraphics[width=8cm]{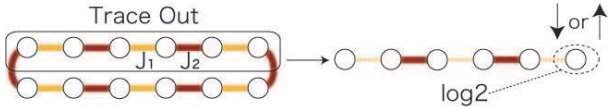}
\end{center}
\caption{(Color Online) Schematic diagram of taking a partial trace in the periodic
 dimerized Heisenberg system. The degrees of freedom are revealed on the
 edges on which the effective $S=1/2$ spins localize.}
\label{fig:PBC-OBC}
\end{figure}

\section{Evaluation of EE for several spin chains}
We calculate the EE for the three spin chains 
and show that the edge state picture of the
 EE is valid for these systems.
We obtain the ground state of the systems numerically using
the Lanczos method in an invariant subspace with $S_{total}^{z}=\sum_i
 S_{i}^z=0$, where $S_{i}^z$ denotes the $z$-component of the spin operator on
the $i$-th site. 
The EE is evaluated by diagonalizing $\hat{\rho}_{A}$ using
the Householder method. We calculate the logarithm in base $2$.  
 We impose periodic boundary conditions for the systems. 
These models exhibit various behaviors that can be distinguished
as different quantum phases by the EE.
\paragraph*{\underline{$S=1/2$ Dimerized Heisenberg Model:}}
The Hamiltonian of the $S=1/2$ dimerized Heisenberg
system is given by
\begin{eqnarray}
H_{S=1/2,D} &=& \sum^{N/2}_{i=1}\left( J_{1} \bm{s}_{2i-1}\cdot \bm{s}_{2i}+
 J_{2} \bm{s}_{2i}\cdot \bm{s}_{2i+1}\right),
\end{eqnarray}
where $\bm{s}_i$ are the $S=1/2$ spin operators. $J_1$ and $J_2$
are parametrized as $J_1=\sin\theta$ and $J_2=\cos\theta$, respectively,
where $\theta\in S^1$.
We consider the case of $-\pi/2<\theta<\pi/2$ in this paper.
The ground state is unique in any $\theta$\cite{Hida}.
In the case of 
$0<\theta\ll\frac{\pi}{4}$, the ground state is composed of an ensemble of
      N/2 singlet pairs and the energy gap is finite
$\left(\sim\frac{3J_1}{2}\right)$. 
The system is equivalent to the isotropic antiferromagnetic
Heisenberg chain at 
$\theta=\frac{\pi}{4}$. This
is the only gapless point.
In the ferromagnetical strong coupling, 
$\theta\rightarrow -\frac{\pi}{2}+0$, the system is
effectively considered as the $S=1$ Heisenberg chains\cite{Hida}.

We now discuss the EE of the system. Fig.\ref{fig:dimerf} shows the
$\theta$ dependence of the EE.
Fig.\ref{fig:PBC-OBC} shows the subsystems considered in this study. As
shown in
Fig.\ref{fig:PBC-OBC}, a partial trace is performed across
two $J_2$ coupling bonds.
The EE tends to diverge 
as the system size increases at the critical point
$\theta=\frac{\pi}{4}$ (isotropic Heisenberg point).
 This behavior of the EE has been discovered by
 Vidal {\it et al.}\cite{Vidal}.
One can clearly observe the EE of the edge states, $2\log 2$,
 near $\theta\sim0$, while the EE
 converges to $0$ in the region of
$\theta\rightarrow \frac{\pi}{2}$.
In the former limit, the system becomes a collection of $N/2$ singlet pairs with two almost
 free spins near the two boundaries.
Therefore, the EE in this region is given by 
that of the two free spins near the boundaries $2\log 2$.
In the latter limit, the system also becomes a collection of
 $N/2$ singlets; however, there are no free spins in this case.
Then, the EE vanishes.

These results can be interpreted as follows. The EE
 reflects the local degrees of freedom
 that appear near the boundaries when the system gets truncated.
Then, we speculate that a lower bound is obtained for EE as ${\cal S}
 \ge\log{g}$,
where $g$ denotes the above mensioned degrees of freedom near the edges. 
In this case,
this local degree of freedom arises due to the two $S=1/2$ spins
localized on the edges as $g=4$ as $|\alpha\rangle_{L}
\otimes |\beta\rangle_{R}$ $(\alpha,\beta=\uparrow,\downarrow)$.

The low energy behavior of this system is identical to
that of the $S=1$ Heisenberg model in the region of 
$\theta\rightarrow -\frac{\pi}{2}+0$\cite{Hida}.
We expect that the edge states of the $S=1$ Haldane phase are similar to those
of the $S=1/2$ dimer phase as the EE is larger than
$2\log{2}$, which gives the lower bound.
\paragraph*{\underline{S=1 XXZ Chain with D-term:}}
\begin{figure}[!tb]
\begin{center}
\includegraphics[width=8cm]{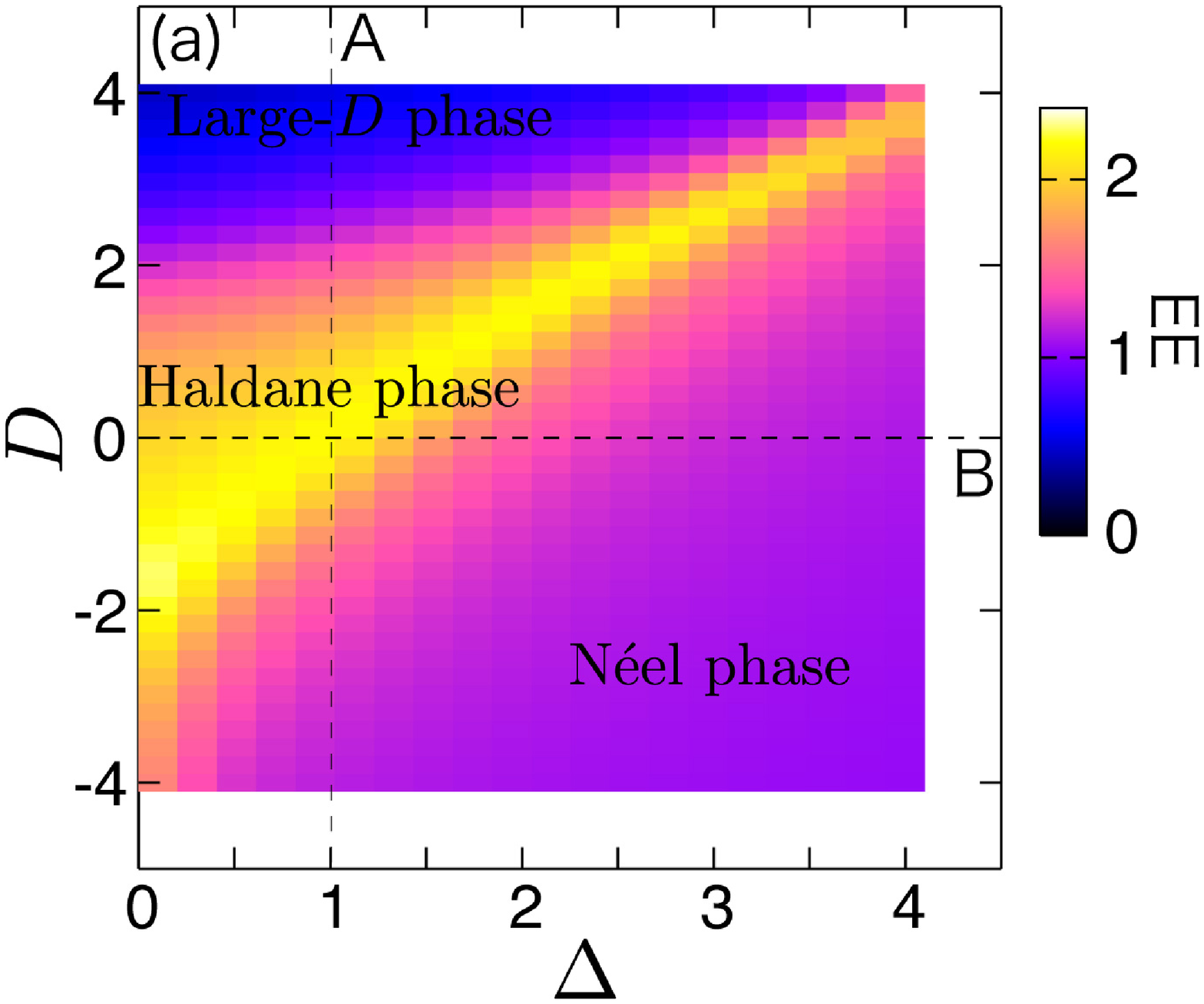}
\end{center}
\begin{center}
\includegraphics[width=6cm]{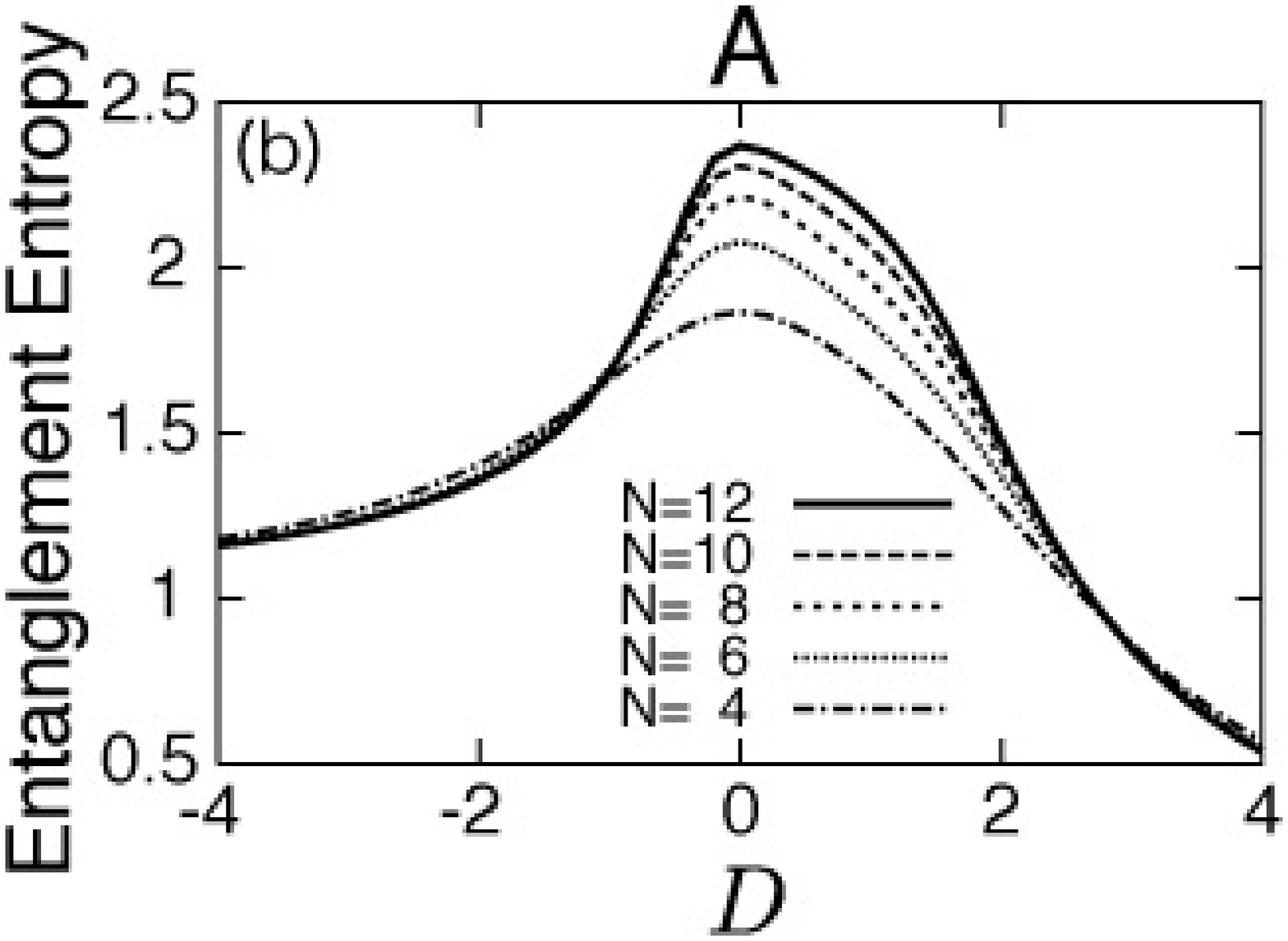}
\end{center}
\begin{center}
\includegraphics[width=6cm]{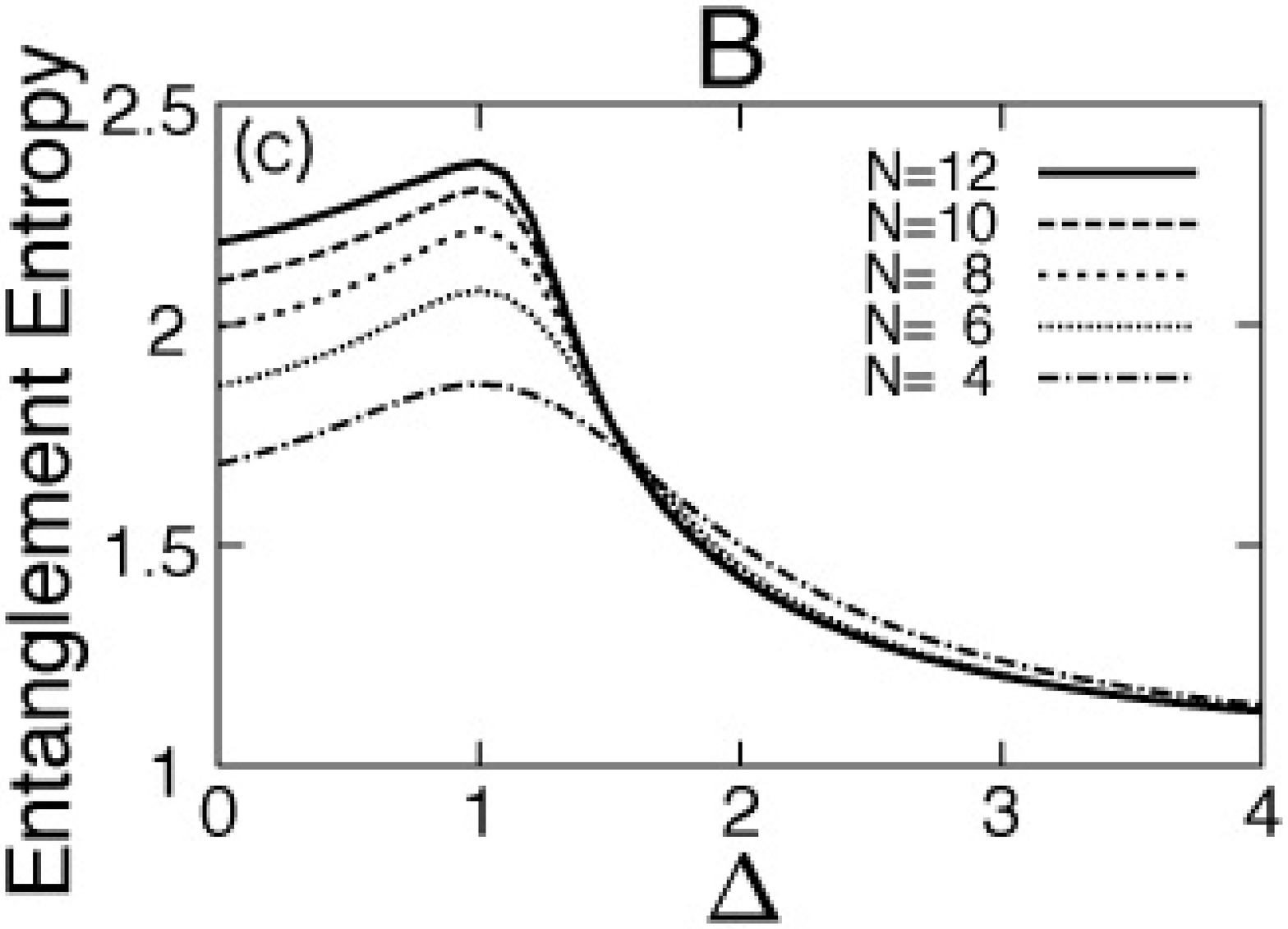}
\end{center}
\caption{(Color Online) (a) The EE of the $S=1$ XXZ chain with the D
 term (N=10). $\Delta$ represents the anisotropies along the
 $S_z$ direction. $D$ is the coefficient of the $D$ term. 
The base of the logarithm is $2$.
(b) Cross section of Fig.\ref{fig:LD8} along line A.
$\Delta=1.0$, N=4,6,8,10,12.(c) Cross section of Fig.\ref{fig:LD8} along
line B. $D=0.0$, N=4,6,8,10,12.}
\label{fig:LD8}
\end{figure}

The Hamiltonian of the $S=1$ XXZ chains with an on-site anisotropic term is given by
\begin{eqnarray}
H_{S = 1} &=& J \sum^{N}_{i = 1} \left( S^{x}_{i} S^{x}_{i+1} + S^{y}_{i}
 S^{y}_{i+1} + \Delta S^{z}_{i} S^{z}_{i+1} \right) 
\nonumber \\& &
 + D \sum^{N}_{i=1} (S^z_i)^2,
\end{eqnarray}
where $\bm{S}_i$ are the $S=1$ spin operators.
The energy gap as a Haldane gap is finite near the point $\Delta=1$ and $D=0$\cite{HaldaneGap}.
There are several phases known as the ``Haldane phase,'' the large-$D$
phase and the N\'{e}el phase in the range of $\Delta >0$ with
a gapped-gapped 
transition (Gaussian transition) between the Haldane phase and the large-$D$
phase and the Ising transition between the N\'{e}el phase and the
Haldane phase. These three phases are distinguished by local and nonlocal 
order parameters\cite{Nijs} or by the breaking of the hidden
$\bm{Z}_2\times\bm{Z}_2$ symmetry
that appears on performing a non-local unitary transformation
\cite{Z2Z2-1,Z2Z2-2}.
Quantum order parameters to characterize the states are also
defined\cite{HatsugaiOrder1,HatsugaiOrder2,HatsugaiOrder3}.
A system with an open boundary condition has the localized
effective spin-$1/2$ edge state in the Haldane
phase\cite{KennedyEdge,Hagiwara}, and the low energy effective
Hamiltonian can be written by an effective coupling of 
the two $S=1/2$ edge spins\cite{KennedyEdge}.

Fig.\ref{fig:LD8}(a) shows the EE of the XXZ
spin chain with the $D$ term, which is calculated in
the subspace $S_z^{total}=0$ by changing the anisotropies $\Delta$ and
the coefficient of the $D$ term $D$ with a system size $N=10$.  
Figs.\ref{fig:LD8}(b) and (c) show the
cross section of Fig.\ref{fig:LD8}(a) along the corresponding planes
 (indicated by lines A and B in Fig.\ref{fig:LD8}(a)) and 
the system size dependence of the EE. 

From Fig.\ref{fig:LD8}(a), it can be observed that there are three regions.
The EEs of the large-$D$ phase, the N\'{e}el
phase, and Haldane phase are approximately $0$, $\log 2$, and $2\log 2$,
respectively.
We now discuss the relation between the EE and each phase.
We use bases $S_i^z|\pm\rangle_i=\pm|\pm\rangle_i$, $S_i^z|0\rangle_i=0$.
The large-$D$ phase (blue region) has a ``quantum-gas-like'' GS.
In the large-$D$ limit, the quantum fluctuation
$\frac{J}{2}\left[S_i^+S_{i+1}^- + S_i^-S_{i+1}^+\right]$ can be ignored.
Then, this state can be approximately described by 
$|\mbox{large-}D\rangle=\bigotimes_{i=1}^{N}|0\rangle_{i}$, which
results in a vanishing EE because this state is the direct product state.
This state does not have any degrees of freedom in the subsystem
created by taking a partial trace.
In the N\'{e}el phase, the GS is N\'{e}el ordered as
$|\mbox{N\'{e}el} \rangle=\bigotimes_{i=1}^{N}|(-1)^i\rangle_{i}$
(quantum-solid-like).
We consider the degrees of freedom of the edge states
created by taking a partial trace.
We denote the spin on the left and right edge as $S_{L}$ and $S_{R}$,
respectively.
Then, $S_L$ is correlated with $S_R$ in this state
due to the N\'{e}el order.
This implies that there are two degrees of freedom. Then, the EE is described
by the effective degrees of freedom as
$|+\rangle_{L}\otimes|-\rangle_{R}$,
$|-\rangle_{L}\otimes|+\rangle_{R}$ ($\mbox{subsystem size}:\mbox{even}$)
and $|-\rangle_{L}\otimes|-\rangle_{R}$,
$|+\rangle_{L}\otimes|+\rangle_{R}$ ($\mbox{subsystem size}:\mbox{odd}$). 
This yields $\log g$ ($g=2$) as the EE.
Finally, the Haldane phase is a quantum liquid. The
states which include $+,-$ particles as many as
the number of $0$ mainly contribute to GS.
There are localized effective spin-$1/2$ edge states in the Haldane
phase.
 The EE is larger than 
$2\log{2}$, which
can be interpreted as the contribution from the edge state. 
The rest of the contribution ${\cal S}-2\log{2}$
comes from the bulk. This interpretation is consistent with the result.
The EE reflects the spontaneous
breaking of the hidden 
$\bm{Z}_2\times\bm{Z}_2$ symmetry,
 which plays an
essential role in Haldane spin chains\cite{Z2Z2-1,Z2Z2-2} and it
 also relates to the edge states.
In summary, the EE has a lower bound of $\log{g}$, where
$g$ is the
number of elements in the group of spontaneously broken symmetry
($g=1$ for the large-$D$ phase, $g=2$ for the N\'{e}el phase, and $g=2^2$ for
the Haldane phase), at least in the calculated regions.

We notice that the EE is lower than $2\log2$ in the Haldane phase
when total system size is small, as shown in figs.3 (b) and (c).
This low value of the EE is due to the correlation length of the system and the
system size.
When the correlation length is long as compared to the total system
size, $S_L$ and $S_R$ are effectively coupled. Then, the
degrees of freedom on the edges decrease because both the sides of the
edge states are not independent in the presence of substantial coupling.

\begin{figure}[!tb]
\begin{center}
\includegraphics[width=
8cm
]{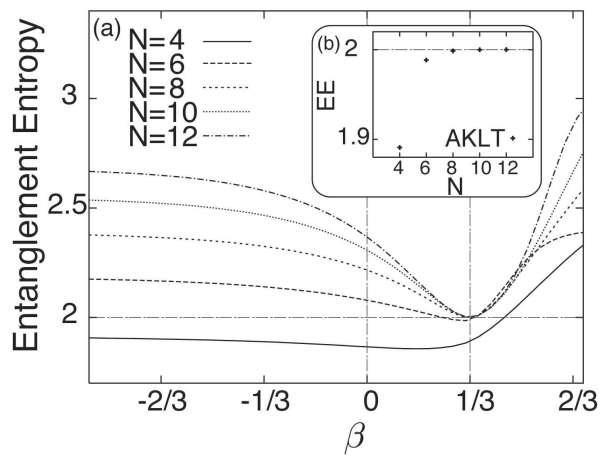}
\end{center}
\caption{(a) The EE of the periodic $S=1$ AKLT model to the Heisenberg
 Chain. (b) Size dependence of the EE at the AKLT point. The base of the
 logarithm is $2$.}
\label{fig:AKLT}
\end{figure}

As discussed above, the bulk-edge correspondence is confirmed
through the trace-out operation, and the EE shows the degrees of freedom
in the subsystem.
An edge state in the open boundary condition is one of the important
characteristics of the Haldane materials. The abovementioned phases of the periodic spin
chains are well characterized by the EE that reflects the edge state of
each corresponding open spin chain.
We found that the bulk and edge are inextricably
linked in these kinds of spin chains.
We confirmed that the Haldane phase can be distinguished by the EE
intermediating the information of the edge states.
\paragraph*{\underline{The $S=1$ Heisenberg Model to the AKLT Model:}}
We also calculate the EE of the AKLT model numerically by
modifying the parameter $\beta$ (Fig.\ref{fig:AKLT}).
The Hamiltonian is given by
\begin{eqnarray}
H_{AKLT}&=& J\sum^{N}_{i=1}\left(  \bm{S}_{i}\cdot \bm{ S}_{i+1}+
 \beta \left(\bm{S}_{i} \cdot\bm{S}_{i+1} \right)^2 \right),
\end{eqnarray}
where $\bm{S}_i$ are the $S=1$ spin operators.
There are solvable points at $\beta=1/3$ \cite{AKLT}and $\beta=\pm 1$\cite{TBpoint1,TBpoint2,ULSpoint1,ULSpoint2,ULSpoint3}.
The ground state can be explicitly written as the valence-bond solid
(VBS) state at the point $\beta=1/3$.

Fig.\ref{fig:AKLT} shows the EE of the AKLT model obtained by changing
the parameter $\beta$.
The result shows that the value of EE is minimum
at the point $\beta=1/3$,
at which the ground state is the VBS state.
The EE is lower than $2\log{2}$ when the system size is
$N=4,6,8,10$. 
This is due to the correlation between the edge spins ($S_L$ and $S_R$), which
is also discussed in the previous model.
According to the edge state picture of the EE, ${\cal S}\ge2\log 2$ when the substantial
coupling of the spins vanishes (in the large $N$ limit), as speculated
from the numerical results;
however, one cannot observe this relation from the numerical calculations.
To confirm this relation, we evaluate the exact form of the EE of the
periodic VBS state analytically.
The EE of the periodic VBS state is obtained
analytically by using a transfer matrix technique\cite{Hirano2} as
follows:
\begin{eqnarray}
{\cal S}_{N,L}&=& -3 \lambda _A \log \lambda _A -   \lambda _B
 \log \lambda _B,\\
\lambda _A &=& \frac{1}{4}\frac{(1-p^L)(1-p^{N-L})}{1-p^{N-1}},\\
\lambda _B &=& \frac{1}{4}\frac{(1+3p^L)(1+3p^{N-L})}{1-p^{N-1}},
\end{eqnarray}
$p=-\frac{1}{3},$
where $N$ and $L$ denote the total system size and the length of
the subsystem, respectively. It is evident that this expression is
invariant under the replacement $L'=N-L$, {\it i.e.}, ${\cal
S}_{N,L}={\cal S}_{N,N-L}$, which is a general property of the EE.

In the limit $N\rightarrow \infty$, the EE reduces to the
expression
\begin{eqnarray}
 \lim _{N\rightarrow \infty}
{\cal S}_{N,L} &=& 2 \log{2} -\frac{3}{4}(1-p^L)\log (1-p^L) \nonumber \\
&&-\frac{1}{4}
(1+3p^L)\log (1+3p^L),
\end{eqnarray}
which is exactly the same as the EE of the VBS state with
two spin-$1/2$s on the boundary\cite{Fan1}.

Next, we consider the thermodynamic limit of ${\cal S}_{N,L}$ with
$\alpha=L/N$ fixed.
We obtain $\lim _{N\rightarrow
 \infty} {\cal S}_{N,N\alpha}=2\log 2$ ($\alpha=L/N$ fixed).
We can see that the EE does not diverge.
This result
is consistent with the equation ${\cal S}\ge 2\log 2$ as 
a consequence of the edge state picture.
According to the edge state picture, 
the EE in the AKLT model receives no contribution from the bulk, while the
contribution from the two edges is present. 
Therefore, this is the point at which the
EE is lowest in the Haldane phase, as long as the edge state picture of
the EE is valid.

\section{Conclusion}
We have explored new aspects of the bulk-edge correspondence in spin
systems on the basis of quantity obtained from the quantum information,
 namely, the entanglement entropy (EE).
The tracing out can be interpreted as
the truncation of the system, and the EE includes the
information regarding how many degrees of freedom the truncated state has.
In the case of the $S=1$ XXZ spin chain with the $D$ term and the AKLT
model, it is
the degrees of freedom of the effective $S=1/2$ spins localized at the edges
of the chain, which are known as the characteristics of the ground state
in the Haldane phase with an open boundary condition. Futher, we speculate
that the EE has a lower bound of $\log g$, reflecting the effective
boundary degrees of freedom $g$.
\begin{acknowledgments}
We wish to thank Y.~Kuge, H.~Song, H.~Katsura, S.~Tanaka, and
 I.~Maruyama for their fruitful suggestions during discussions.
The program used in this study to diagonalize the spin chains
is based on the package TITPACK ver.2.
We also used LAPACK to diagonalize
the matrices.
The computation in this work has been done using the
facilities of the Supercomputer Center, Institute for Solid
State Physics, University of Tokyo.
YH was supported by Grants-in-Aid for Scientific Research from JSPS
(No. 17540347), from Scientific Research
on Priority Areas from MEXT (No.18043007), and from the Sumitomo Foundation. 
\end{acknowledgments}

\end{document}